\documentclass{ws-procs9x6}

\begin{document}

\title{Highlights on beauty detection in nucleus-nucleus collisions with ALICE}

\author{Rosario Turrisi}
\author{\em for the ALICE collaboration}

\address{ Universit\`a degli Studi e INFN -- Padova \\
          via Marzolo, 8 - I35131 Padova, Italy }

\maketitle

\abstracts{
We describe a strategy for the detection of {\em open beauty} 
in the semi-electronic channel with ALICE evaluating the expected S/(S+B) ratio.
}
Heavy-ion physics at the LHC, where Pb-Pb collisions 
at 5.5A TeV will be produced, is focussed on the study of the behavior of strongly
interacting matter at extreme energy density over a large volume, 
where the formation of the Quark Gluon Plasma (QGP) is expected.
Beauty is, in this context, an especially interesting observable. The $b$-quark energy loss,
which can be studied by comparing the $p_T$ spectra from AA and pp collisions, 
can provide information on the composition of the medium.
The measurement of the heavy quark cross sections will provide the natural 
normalization for the analysis of quarkonia suppression, one of the main signals of QGP formation and
the amount of $b \to \mathrm{J}/\psi$ contamination to the direct J/$\psi$ yield.
Beauty hadrons can be detected through their semi-leptonic decays.
In ALICE\cite{tp}, an LHC experiment dedicated to heavy-ion physics,
this has been studied for the semi-electronic channel.
\\
The ALICE detector is designed to handle a multiplicity currently estimated as 
4000-8000 charged particles per rapidity unit at midrapidity per event.
Among the many detectors which make up ALICE, three are functional for this study.
The Time Projection Chamber\cite{tpctdr} (TPC)
and the Inner Tracking System\cite{itstdr} (ITS) provide
tracking and vertexing, at $p_T$ as low as 200~MeV with a magnetic field $B=0.4$ T.
The Transition Radiation Detector\cite{trdtdr} (TRD) allows electron/hadron separation.
The TPC can also be used for particle identification purposes.
\\
Electrons coming from $b$ weak decays are characterized by hard $p_T$ spectrum and
average impact parameter $<d_0>$ of the order of a few hundred microns.
The $d_0$, defined as the distance of closest approach of the 
particle trajectory to the interaction vertex in the plane orthogonal to the beam,
is measured by the ITS, which is equipped with a two-layers Silicon Pixel Detector.
The resolution is $\sigma_{d0} \lesssim 70 \mathrm{\mu m}$ at $p_T = 1 \mathrm{GeV}$.
\\
The simulation, performed with the AliRoot\cite{aliroot} package, 
includes all the main sources of background electrons: decays of prompt charm,
pair production due to photon conversion in the detector materials,
Dalitz decays of light and strange mesons, pions misidentified as electrons.
\\
The signal, $b \to e + X$, and the $c \to e + X$ 
background have been generated using PYTHIA6\cite{pythia}.
Background electrons and charged pions have been produced by HIJING\cite{hijing}, 
tuned to reproduce a charged particles density of $dN_{ch}/dy = 6000$ at midrapidity.
Beauty and charm cross-sections are fixed to 1.79 and 45 barn respectively, which is our
extrapolation of NLO calculations\cite{mnr} to Pb-Pb at 5.5 A TeV.
\\
Particle transport is done by GEANT3 in the frame of the detailed detector geometry as implemented in AliRoot. 
Tracking inefficiency in the TPC has been accounted for
via a parameterization\cite{paratpc}, while for the ITS, which determines the ALICE vertexing performance,
the AliRoot Kalman-filter-based algorithm has been used.
\\
Electron identification relies on both the TRD and the TPC. The TRD allows a pion rejection 
factor whose value depends on the electron identification efficiency, for $p_T > 1 $GeV. 
We have chosen the combination $\epsilon_e=0.9, \epsilon_\pi=0.01$ 
($1/100$ of pions misidentified as electrons) which provides a sample purity 
of $\sim 50 \%$. The contribution of heavier hadrons after TRD filtering is assumed to be negligible. 
We have then extended the identification power by using the 
specific energy-loss in the TPC. This technique allows, in the momentum range relevant to this
study, to obtain a combined pion rejection factor of $10,000$.
All other background sources are reduced by cutting on $d_0$.
\begin{figure}[ht]
\centerline{\epsfxsize=2.66in\epsfbox{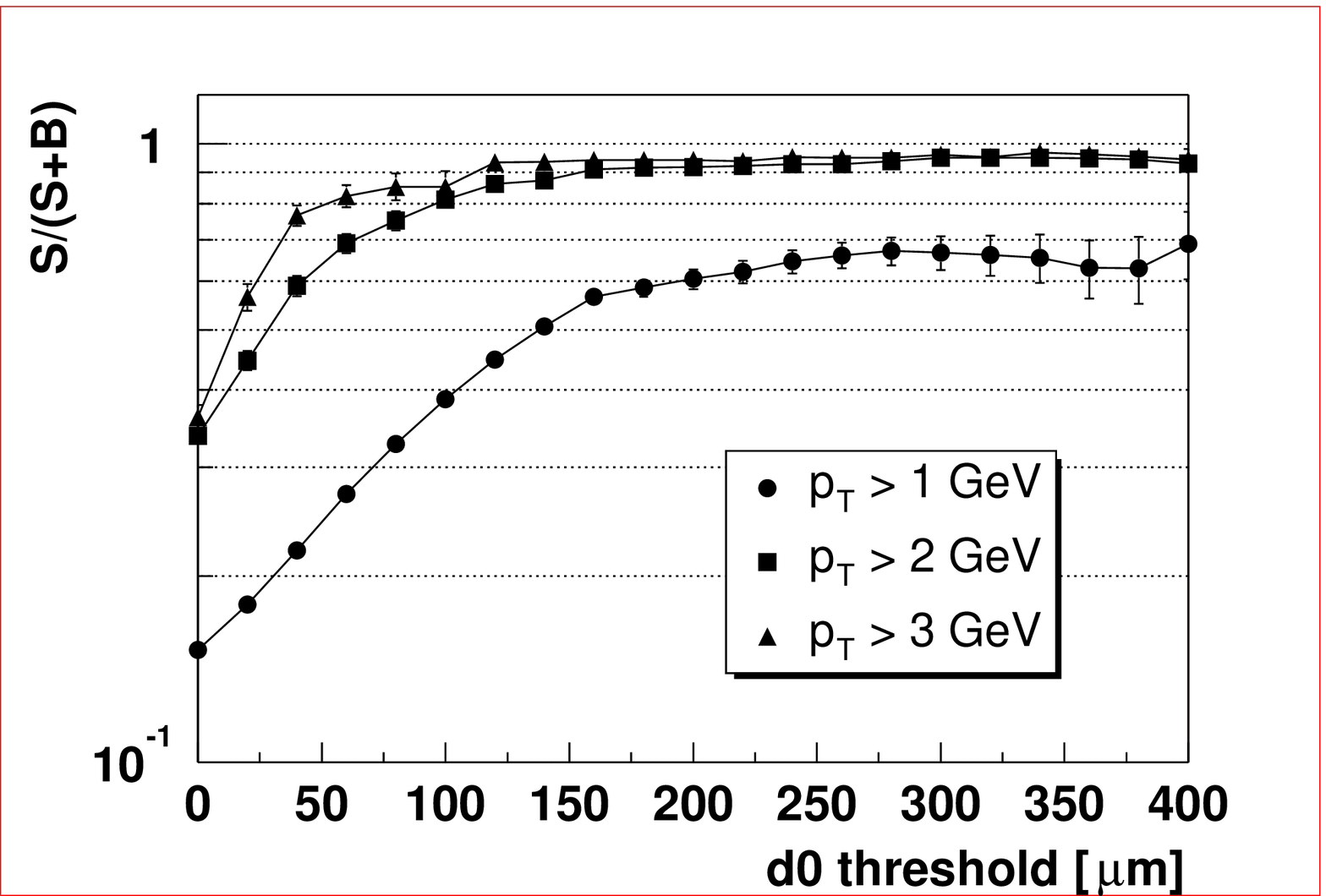}\epsfxsize=2.66in\epsfbox{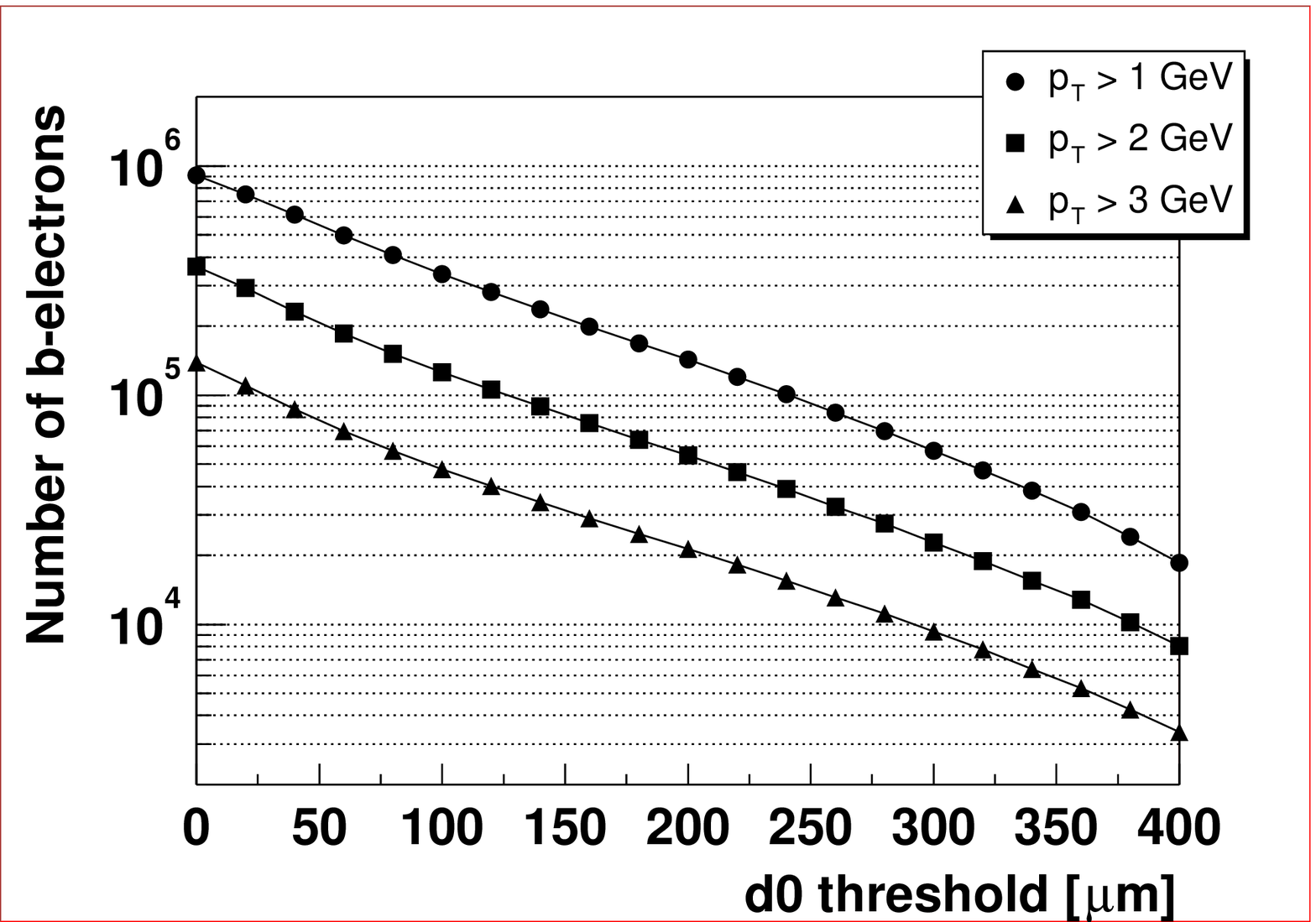}}   
\caption{S/(S+B) (left) and detected $b$ electrons (right) as a function 
of d$_0$/p$_T$ threshold.\label{sovnp}}
\end{figure}
In Fig.\ref{sovnp}, left panel, we show the S/(S+B) ratio for p$_T >$ 1, 2 and 3~GeV
as a function of the lower threshold on $d_0$.
As an example, a condition
of $p_T > 2 $GeV and $180 \leq d_0 \leq 500 \mu$m assures a 90\% purity.
The upper cut on $d_0$ is applied to reduce long-lived background and 
tracks which suffered large-angle scattering in the detector material.
Finally, this detection strategy yields, given the above hypotheses on cross-sections
and background level, a number of beauty semi-electronic decays $N\simeq 8\times 10^4$ in $10^7$ 
central Pb-Pb events, corresponding to one-month run of ALICE. 
A view of the attainable statistics of b-hadrons
is shown in Fig.\ref{sovnp} on the right panel as a function of the $d_0$ threshold for three
$p_T$ ranges. 
We assumed 100\% tracking efficiency in the TRD. This parameter is currently under study:
it will scale the statistics by a multiplicative factor, while will not change the S/(S+B).
Additional statistics will be provided by the $b \to c + X \to e + X$ decays, estimated to be 
$\sim 15\%$ of the direct beauty electrons at $p_T = 2 $GeV and not included in the present study for technical reasons.
\\
In conclusion this study provides a positive indication on the capability of ALICE to 
select a clean and high statistics sample of {\em open beauty}. Further study 
is in progress to quantify the accuracy on the cross section and 
on the transverse momentum spectrum shape.
\\

\end{document}